\documentclass[12pt]{article}

\usepackage{preprintstyle}
\usepackage[utf8]{inputenc}
\usepackage{comment}
\usepackage{tikz}

\newcommand{\beq}{\begin{equation}}
\newcommand{\eeq}{\end{equation}}
\newcommand{\bal}{\begin{aligned}}
\newcommand{\eal}{\end{aligned}}
\newcommand{\bc}{\begin{cases}}
\newcommand{\ec}{\end{cases}}
\newcommand{\rmd}{\mathrm d}

\usepackage{dsfont}
\usepackage[normalem]{ulem}
\usepackage{cancel,xcolor}

\title{Complex Geodesics in the Nariai Geometry}

\author{Lars Aalsma$^a$,}
\author{Mir Mehedi Faruk$^b$}
\emailAdd{laalsma@d.umn.edu}
\emailAdd{muturza3.1416@gmail.com}
\affiliation{$^a$Department of Physics and Astronomy, University of Minnesota Duluth, 1049 University Drive, Duluth, MN 55812, USA}
\affiliation{$^b$Center for Fundamental Physics, School of Physical Science and Technology,
ShanghaiTech University, 393 Middle Huaxia Road, Shanghai 201210, China}

\abstract{We study two-point correlation functions of heavy scalar fields in the Nariai geometry. Utilizing the heat kernel formalism, we obtain this result from a geodesic approximation to the two-point function on a product of spheres. By analytically continuing one of the spheres, we obtain the correlation function in the Nariai geometry. This result involves a sum over complex geodesics, extending previous results in pure de Sitter space. We emphasize the important role of the phase of each geodesic contribution, which needs to be taken into account to avoid spurious singularities in the correlator.}

\begin{document}

\maketitle

\section{Introduction}
Geodesics play an important role in studying semiclassical aspects of gravity, as well as in quantum gravity. One tool in particular that has been useful is the geodesic approximation to correlation functions. For sufficiently heavy fields, the path integral used to compute two-point functions localizes to the classical worldline that connects the two points. Also taking into account the one-loop correction describing (quantum) fluctuations away from this classical trajectory gives an accurate approximation to the exact result. Applications in the context of AdS/CFT include \cite{Fidkowski:2003nf,Brecher:2004gn,Festuccia:2005pi,Ceplak:2024bja,Afkhami-Jeddi:2025wra}, where features of the black hole interior are captured by geodesics that connect points outside the horizon and approach the black hole singularity. Interestingly, certain features of this correlator seems to be captured by complex geodesics.

Complex geodesics appear naturally when studying de Sitter space, as not all points in opposite static patches can be connected by a real geodesic. Nonetheless, it has been demonstrated in \cite{Aalsma:2022eru,Chapman:2022mqd} (see also \cite{Fischetti:2014uxa}), that complex geodesics still can be used to give an approximation to the Green's function in de Sitter space. Applications of these complex geodesics (and surfaces) appear in the context of timelike entanglement entropy, see e.g. \cite{Doi:2023zaf}.

In this paper, we extend our earlier work on complex geodesics in pure de Sitter space to the near-horizon geometry of Nariai black holes. Part of the motivation for this work is to extend the geodesic approximation in de Sitter space to geometries that include black holes. In particular, non-perturbative effects such as black hole nucleation have been suggested to play a role in a de Sitter version of Maldacena's information paradox \cite{Aalsma:2022eru,Mirbabayi:2023vgl}, so it is of interest to see how the geodesic approximation is modified compared to pure de Sitter space.

To this end, we compute the Green's function of scalar fields on a product of spheres. After taking a large-mass limit to reveal a geodesic approximation we analytically continue one of the spheres to de Sitter space to obtain the geodesic approximation in the Nariai geometry. Because we keep the radii of the spheres distinct, our results are applicable to a large class of near-horizon Nariai limits. 

In Section \ref{sec:SphereGreen}, we discuss the Green's function on a (product of) sphere(s) and take a large-mass limit to reveal the geodesic approximation. Special attention is paid to the phase of the different contributions, which are important to ensure that the result is real and does not contain any spurious singularities. In Section \ref{sec:Nariai} we discuss the analytical continuation necessary to obtain the Green's function in de Sitter space. We combine these results to obtain the Green's function in the Nariai geometry. We conclude with a discussion in \ref{sec:discussion}. Useful mathematical identities and details regarding the coordinate systems used throughout this work are provided in the appendices.

\section{Green's functions on spheres} \label{sec:SphereGreen}

\subsection{Heat kernel techniques}
To compute the Green's function on a general background, heat kernel techniques are often useful (see \cite{Vassilevich:2003xt} for a review). For a general differential operator ${\cal O}(x)$ in $d$ dimensions, the Euclidean Green's function $G_E(x,y)$ obeys,
\beq
{\cal O}(x)G_E(x,y) = \frac{\delta^{d}(x-y)}{\sqrt{g(x)}} ~.
\eeq
Here, $g(x)$ is the metric determinant. Using the Schwinger-de Witter proper-time formalism, the Green's function can be expressed in terms of the so-called heat kernel $K_E(x,y;s)$, which solves the heat equation given by
\beq
(\partial_s+{\cal O})K_E(x,y;s) = 0 ~,
\eeq
with boundary condition
\beq
\lim_{s\to 0}K_E(x,y;s)=\frac{\delta^{d}(x-y)}{\sqrt{g(x)}} ~.
\eeq
The relation between the Green's function and heat kernel is
\beq
G_E(x,y) = \int_0^{\infty}\rmd s \,K_E(x,y;s) ~.
\eeq
A useful representation of the heat kernel can be found by expanding it into eigenfunctions of the operator ${\cal O}$. Assuming discreteness of the spectrum, we denote an eigenfunction by $\Psi_l(x)$ where $l$ labels different eigenvalues. We can then write,
\beq
K(x,y;s) = \sum_l e^{-\lambda_l s}\Psi_l(x)\Psi_l(y) ~,
\eeq
where $\lambda_l$ are the eigenvalues. The eigenfunctions are normalized as,
\begin{align}
\sum_l \Psi_l(x)\Psi_l(y) &= \frac{\delta^{d}(x-y)}{\sqrt{g(x)}} ~, \label{eq:Norm1}\\
\int\rmd^{d}x\sqrt{g}\,\Psi_l(x)\Psi_{l'}(x) &= \delta_{ll'} ~.  \label{eq:Norm2}
\end{align}
One of the advantages of working with the heat kernel is that on product spaces the heat kernel factorizes whereas the Green's function in general does not.

\subsection{Green's function on $S^2$}
We now study the Green's function of a massive scalar field on an $S^2$. The operator of interest is then given by
\beq
{\cal O} = -\nabla^2+m^2 ~.
\eeq
Here $\nabla^2$ is the Laplacian on the $S^2$ and $m$ is the mass of the scalar field. The explicit coordinates and metric we'll use to describe the $S^2$ are
\beq
\rmd s^2 = \ell^2(d\theta^2+\sin^2\theta\rmd\phi^2) ~.
\eeq
The eigenfunctions of the Laplacian are spherical harmonics, which can be expanded in terms of Legendre polynomials. Due to coordinate invariance, the Green's function can only depend on the proper distance on the $S^2$. We can always choose a frame such that the separation in the angle $\phi$ is zero such that the proper distance is $D = \ell\Theta$, where we distinguish the coordinate $\theta$ from the distance $\Theta = \theta_2-\theta_1$ between two points. It is then convenient to use the notation $p=\cos\Theta$. Without loss of generality, we can therefore write a product of eigenfunctions as $\Psi_l(x)\Psi_l(y)=\Psi_l(p)\Psi_l(1)$. Concretely, the eigenfunction can be expressed in terms of a Legendre polynomial $P_l(p)$ as
\beq
\Psi_l(p)=\sqrt{\frac{2l+1}{4\pi \ell^2}}P_l(p) ~,
\eeq
where the prefactor is fixed by the normalization conditions. It is now straightforward to check that
\beq
{\cal O}\,\Psi_l(p) = \left(\frac{l(l+1)}{\ell^2} + m^2\right)\Psi_l(p) ~.
\eeq
The heat kernel therefore takes the form
\beq
\bal
K_{S^2}(p;s) &= \sum_{l=0}^\infty e^{-s\left(\frac{l(l+1)}{\ell^2}+m^2\right)}\Psi_l(p)\Psi_l(1) \\
&=\frac{1}{4\pi\ell^2}\sum_{l=0}^\infty(2l+1) e^{-s\left(\frac{l(l+1)}{\ell^2}+m^2\right)}P_l(p) ~.
\eal
\eeq
The factor of $2l+1$ corresponds to the degeneracy of an eigenvalue at level $l$. Integrating over $s$ yields the Green's function
\beq
G_{S^2}(p) = \frac1{4\pi}\sum_{l=0}^\infty \frac{2l+1}{l(l+1)+m^2\ell^2}P_l(p) ~.
\eeq
It is convenient to express the mass in terms of $\Delta_\pm$, defined by
\beq
\Delta_\pm = \frac12\left(1\pm \sqrt{1-4m^2\ell^2}\right) ~.
\eeq
Using the identity \eqref{eq:HyperLegendre}, the sum can be expressed in terms of a hypergeometric function
\beq \label{eq:ExactS2Green}
G_{S^2}(\Theta) =  \frac{\Gamma(\Delta_+)\Gamma(\Delta_-)}{4\pi}\,_2F_1\left(\Delta_+,\Delta_-,1,\frac{1+\cos\Theta}{2}\right)~.
\eeq
We now show that, in a large-mass expansion, the Green's function takes a simple form in terms of the geodesic that connects the two points.

To see this, we write the hypergeometric function in terms of a Legendre polynomial (using \eqref{eq:LegendreAsHyper}) such that
\beq
\,_2F_1\left(\Delta_+,\Delta_-,1,\frac{1+\cos\Theta}{2}\right) = P_{-\Delta_+}(-\cos\Theta) ~.
\eeq
In the large-mass limit $\Delta_\pm = \frac12\pm im\ell$. Then, using the asymptotic expansion \eqref{eq:asympLegendre} we obtain 
\beq \label{eq:ExactExpansion}
\,_2F_1\left(\Delta_+,\Delta_-,1,\frac{1+\cos\Theta}{2}\right) = \frac{e^{-m\ell\Theta}}{\sqrt{8\pi m\ell \sin\Theta}}\left(1+{\cal O}(m\ell)^{-1}\right) \qquad (0<\Theta< \pi) ~,
\eeq
which is valid only for $0<\Theta< \pi$, as there is a singularity at $\Theta=\pi$ that is absent in the exact result \eqref{eq:ExactS2Green}.

\subsubsection*{Geodesic approximation on $S^2$}
To derive the Green's function, we used a representation of the heat kernel in terms of eigenvalues of the Laplacian. However, as was reviewed in \cite{Camporesi:1990wm} a more geometric representation exists where the heat kernel can be expressed as a sum over geodesic paths. Explicitly, for a $d$-dimensional space this representation is given by \cite{Vassilevich:2003xt}
\beq \label{eq:GeneralSmalls}
K(x,y;s) = \sum_{\rm geodesics}\frac{\Delta(x,y)^{1/2}}{(4\pi s)^{d/2}}e^{-m^2s-\frac{D^2}{4s}}\sum_{j=0}^\infty a_j(x,y)s^j ~.
\eeq
Here $\Delta(x,y)$ is the Van Vleck-Morette determinant, $D(x,y)$ is the geodesic distance between the points $x$ and $y$ and the expansion in terms of coefficients $a_n$ is known as the heat kernel expansion. Since it describes an expansion around flat space \cite{Vassilevich:2003xt}, the leading coefficient is given by $a_0 = 1$. The sum over geodesics appearing in the front is especially important for compact spaces, since, in principle, there are an infinite amount of geodesics that connect two points and care needs to be taken to identify the ones that give the dominant contribution. Due to the exponential dependence the leading contribution is typically given by the path with the shortest geodesic distance. For symmetric spaces, such as $S^2$, the exact result for the heat kernel can be expressed as an infinite sum over geodesics. For an $S^2$ with radius $\ell$ the exact heat kernel in the geodesic representation is given by \cite{Camporesi:1990wm}
\beq \label{eq:CamporesiExact}
K_{S^2}(\Theta;s) = \frac{\sqrt{2}\ell e^{-m^2s+\frac{s}{4\ell^2}}}{(4\pi s)^{3/2}}\sum_{n=-\infty}^{+\infty}(-1)^n\int_\Theta^\pi \rmd\Phi\frac{(\Phi+2\pi n)}{\sqrt{\cos\Theta-\cos\Phi}}e^{-\frac{\ell^2(\Phi+2\pi n)^2}{4s}} ~.
\eeq
Here, the different geodesics connecting the two points are labeled by $n$. Importantly, geodesics with even $n$ or odd $n$ pick up a different phase, which is captured by the factor $(-1)^n$. The phase of the different geodesic contributions is important to ensure that the result for the geodesic approximation to the Green's function is real and finite in the limit $\Theta\to \pi$, as we expect from \eqref{eq:ExactS2Green}.

For separations $0<\Theta<\pi$, geodesics with $n\neq 0$ are exponentially suppressed. Focusing on the term with $n=0$, we can obtain a small-$s$ expansion and see how that compares with the general expression \eqref{eq:GeneralSmalls}. This case, corresponds to the shortest path connecting the two points. We will now show how to extract the leading behavior from \eqref{eq:CamporesiExact} in this limit. We want to obtain the leading contribution to the integral
\beq
I(\Phi,s) = \int_\Theta^\pi\rmd\Phi \frac{\Phi e^{-\frac{\ell^2\Phi^2}{4s}}}{\sqrt{\cos\Theta-\cos\Phi}} ~,
\eeq
in the limit $s\to 0$. Since $\Phi\in(\Theta,\pi)$ with $\Theta\in(0,\pi)$ we see that for small $s$, the integral is dominated by the lower boundary of $\Phi$ due to the exponential suppression. We therefore rewrite the integrand in terms of $\Phi = \Theta+x$, where $0<x\ll 1$. In the exponent, we keep the term linear in $x$, since $x/s$ is not necessarily small, but drop terms higher order in $x$. The integral then become
\beq
\lim_{s\to 0}I(x,s) = \int_0^{\pi-\Theta}\rmd x \frac{\Theta}{\sqrt{x\sin\Theta}}e^{-\frac{\ell^2}{4s}\Theta(\Theta+2x)} ~.
\eeq
Lastly, because the integral is dominated by small $x$, we can extend the upper boundary to infinity as the error we make by doing so is exponentially suppressed in the limit $s\to 0$. We note that this is only justified when $\Theta< \pi$. Thus, we obtain
\beq
\lim_{s\to 0}I(x,s) = \int_0^\infty\rmd x \frac{\Theta}{\sqrt{x\sin\Theta}}e^{-\frac{\ell^2}{4s}\Theta(\Theta+2x)} = \sqrt{\frac{2\pi s\,\Theta}{\ell^2\sin\Theta}}e^{-\frac{\ell^2\Theta^2}{4s}} ~.
\eeq
Using this result, we find that the leading small-$s$ behavior of the heat kernel is
\beq
K_{S^2}(\Theta;s) = \frac{e^{-m^2s+\frac{s}{4\ell^2}}}{4\pi s}\sqrt{\frac{\Theta}{\sin\Theta}} e^{-\frac{\ell^2}{4s}\Theta^2}\left(1+{\cal O}(s)\right) ~.
\eeq
Comparing with \eqref{eq:GeneralSmalls}, we see that we can identify the Van Vleck-Morette determinant for an $S^2$ as
\beq
\Delta(\Theta)^{1/2} = \sqrt{\frac\Theta{\sin\Theta}} ~.
\eeq
Integrating over $s$ to obtain the Green's function and expanding for $m\ell \gg 1$ we find
\beq \label{eq:SingleGeodSphere}
G_{S^2}(D) = \frac{1}{\sqrt{8\pi m \ell \sin(D/\ell)}}e^{-mD}\left(1+{\cal O}(m\ell)^{-1}\right)  \qquad (0<\Theta< \pi) ~,
\eeq
where we expressed the result in terms of the geodesic distance $D=\ell\Theta$.

Due to the form of the Van Vleck-Morette determinant, this result is singular for a separation $\Theta=\pi$, which is outside of the regime of validity of the single-geodesic approximation. In particular, the dominant additional geodesic that we need to take into account connects two points by going ``the other way around'' the sphere, leading to a geodesic distance of $\bar  D=\ell(2\pi-\Theta)$. Although this geodesic is subdominant for separations $0<\Theta< \pi$, it gives an equal contribution to the Green's function precisely when $\Theta = \pi$, where the single-geodesic result becomes singular.

\begin{figure}[t]
\centering
\includegraphics{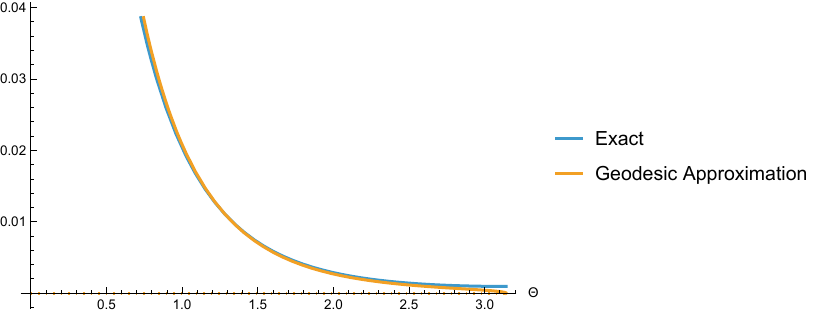}
\caption{Comparison of the exact Green's function \eqref{eq:ExactS2Green} and the geodesic approximation \eqref{eq:SphereGeod}. We took $m\ell = 2$.}
\label{fig:SphereGeo}
\end{figure}

As explained in \cite{Camporesi:1990wm} (see also \cite{Schulman:2012}) the contribution of this second geodesic can be taken into account, simply by replacing the geodesic distance $D$ with the distance of the indirect geodesic $\bar D$ and by taking into account the phase $e^{-i\frac\pi 2}$ that is picked up when crossing the singular point $\Theta=\pi$. Adding both contributions, we obtain the geodesic approximation in the large-mass limit taking into account the leading correction due to the indirect geodesic,
\beq \label{eq:S2approxInter}
G_{S^2}(D) = \frac{1}{\sqrt{8\pi m \ell \sin(D/\ell)}}e^{-mD}+\frac{e^{-i\frac\pi 2}}{\sqrt{8\pi m \ell \sin(\bar D/\ell)}}e^{-m\bar D} + {\cal O}(m\ell)^{-1} ~.
\eeq
This can be further simplified to
\beq \label{eq:SphereGeod}
G_{S^2}(D) =  \frac{e^{-m\ell\pi}\sinh\left(m\frac{\bar D-D}{2}\right)}{\sqrt{2\pi m \ell \sin(D/\ell)}}\left(1+{\cal O}(m\ell)^{-1}\right) ~.
\eeq
We note that the phase of the indirect geodesic is such that it gives a real contribution with a minus sign to the Green's function, consistent with the factor $(-1)^n$ in \eqref{eq:CamporesiExact}, as the indirect geodesic is parametrized by $n=-1$. The result \eqref{eq:SphereGeod} approximates the exact result well, see Figure \ref{fig:SphereGeo}.

\subsection{Green's function on $S^2\times S^2$}
We now turn our attention to a direct product of two $S^2$'s, which is the Euclidean continuation of the near-horizon limit of the four-dimensional Nariai geometry we will study later. The metric we will take is
\beq
\rmd s^2 = L^2(d\psi^2+\sin^2\psi\rmd\chi^2) +  \ell^2(d\theta^2+\sin^2\theta\rmd\phi^2)~.
\eeq
On a product space, the Laplacian factorizes into a sum of the Laplacians of each factor. As such, it can be seen from the heat equation that the heat kernel factorizes in terms of a product of each factor. We can therefore simply use our result for a single $S^2$ to write the combined heat kernel as
\beq
K_{S^2\times S^2}(p,q;s) = \frac{1}{(4\pi L\ell)^2}\sum_{l,k=0}^\infty(2l+1)(2k+1) e^{-s\left(\frac{l(l+1)}{L^2}+\frac{k(k+1)}{\ell^2}+m^2\right)}P_l(p)P_k(q) ~,
\eeq
where now $(p,q)=(\cos\Psi,\cos\Theta)$ and $\Psi=\psi_2-\psi_1$ is the angular separation on the first $S^2$ with radius $L$.

In principal, one can now integrate over $s$ to obtain the exact Green's function
\beq
G_{S^2\times S^2}(p,q) = \frac1{16\pi^2}\sum_{l,k=0}^\infty\frac{(2l+1)(2k+1)}{l(l+1)\ell^2+k(k+1)L^2+m^2L^2\ell^2}P_l(p)P_k(q) ~.
\eeq
We note that the denominator involves terms that couple the radius of one of the $S^2$ factors to the eigenvalues of the other $S^2$. This is a reflection of the fact that, unlike the heat kernel, the Green's function does not factorize into a product of the individual Green's functions of each of the $S^2$ factors.\footnote{This is expected, as we are essentially computing a functional determinant of the sum of two operators which, in general, cannot be written in terms of the determinants of the individual operators.}

However, because the heat kernel does factorize we can focus on the dominant (small-$s$) behavior in both heat kernels. The heat kernel on $S^2\times S^2$ is then simply a product of the individual small $s$ heat kernels
\beq
K_{S^2\times S^2}(D_1,D_2;s) = \sum_{\rm geodesics}\frac{e^{-m^2s+\frac{s}{4L^2}+\frac{s}{4\ell^2}}}{(4\pi s)^2}\sqrt{\frac{(D_1/L)(D_2/\ell)}{\sin(D_1/L)\sin(D_2/\ell)}} e^{-\frac{D_1^2+D_2^2}{4s}}\left(1+{\cal O}(s)\right) ~.
\eeq
The two geodesic distances are defined as $D_{i=1,2} = (L\Psi,\ell\Theta)$ and the geodesic distance on the product space is $D_{S^2\times S^2} = \sqrt{D_1^2+D_2^2}$.

Integrating over $s$ to obtain the Green's function, we now get a sum of four distinct contributions, labeled by two geodesics for each $S^2$. This yields
\beq \label{eq:S2S2Green}
G_{S^2\times S^2}(D_1,D_2) = {\cal G}_{1,2}+e^{i\frac\pi 2}{\cal G}_{\bar 1,2}+e^{i\frac\pi 2}{\cal G}_{1,\bar2}+{\cal G}_{\bar1,\bar2} ~.
\eeq
The four different contributions are labeled by the two possible (dominant) geodesics on each of the $S^2$ factors. Explicitly,
\beq \label{eq:S2S2terms}
{\cal G}_{i,j} = \sqrt{\frac{m(D_i/\ell_i)(D_j/\ell_j)}{32\pi^3\sin(D_i/\ell_i)\sin(D_j/\ell_j)(D_i^2+D_j^2)^{3/2}}}e^{-m\sqrt{D_i^2+D_j^2}} ~.
\eeq
where a bar denotes the conjugate geodesic obtained by sending $D_i\to 2\pi \ell_i -D_i$. The phases in front of the different contributions are chosen such they correctly take into account the factor $(-1)^{n_1+n_2}$ in front of the exact result of the heat kernel. In addition, this ensures that the result does not diverge when either separation $\Psi$ or $\Theta$ equals $\pi$ (although these limits are outside the regime of validity of the geodesic approximation).

\section{Complex geodesics in Nariai black holes} \label{sec:Nariai}
Now that we have determined the Green's function on $S^2\times S^2$ using a geodesic approximation, we will analytically continue one of the spheres to dS$_2$, to obtain the Green's function on the dS$_2\times S^2$ geometry. We will first show how to obtain the dS$_2\times S^2$ geometry as the near-horizon Nariai limit of a charged black hole in de Sitter space.

\subsection{The Nariai geometry}
Our focus will be on taking the near-horizon Nariai limit of four-dimensional (electrically) charged black holes in de Sitter space. However, the final dS$_2\times S^2$ geometry is universal in the sense that it also arises by taking the same Nariai limit in other de Sitter black hole solutions, such as Kerr-de Sitter. See \cite{Mariani:2025hee} for a recent discussion of this limit for instance. From the point of view of the metric, the only difference between different Nariai solutions is the ratio of the radius of the two-dimensional de Sitter space and the radius of the two-sphere.\footnote{This is only true if we just focus on the metric. If we include perturbations of other matter fields in the near-horizon limit, the different near-horizon Nariai theories do become distinct due to their different field content.}

The Reissner-Nordström de Sitter black hole solution can be found in many references (see e.g. \cite{Romans:1991nq} for early discussions and references therein). Here, we just highlight the most important aspects. The metric and gauge field are given by
\beq
\bal 
\rmd s^2 &= -f(r)\rmd t^2 + f(r)^{-1}\rmd r^2 + r^2\rmd \Omega_2^2 ~,\\
f(r)&= -\frac{(r-r_a) (r-r_b) (r-r_c) (r+r_a+r_b+r_c)}{L_4^2 r^2}  ~, \\
A&= -\frac{Q}{4\pi r}\rmd t ~.
\eal
\eeq
Here $r_a$ is the inner black hole horizon, $r_b$ the outer black hole horizon and $r_c$ the cosmological horizon. They are tied together by the four-dimensional de Sitter radius $L_4$ (distinguished from the two-dimensional radius $L$ by the subscript) as
\beq
L_4^2 = r_a^2+r_b^2+r_c^2 + r_ar_b+r_ar_c+r_br_c ~.
\eeq
The relation between the radii and the mass and charge is
\beq
\bal
M & = \frac{1}{2G_4}\frac{r_b^2(r_b^2-L_4^2)-r_c^2(r_c^2-L_4^2)}{(r_c-r_b)L_4^2} ~,\\
Q^2 &=  \frac{4\pi}{G_4}\frac{r_c(r_c^2-L_4^2)-r_b(r_b^2-L_4^2)}{(r_c^{-1}-r_b^{-1})\ell^2_4} ~.
\eal
\eeq
A convenient way to take the near-horizon Nariai limit was discussed in \cite{Aalsma:2025lcb}. The idea is to identify the radial location of the unique non-accelerating observer located in between the outer black hole horizon and cosmological horizon. This location is determined by $f'(r) = 0$ and denoted by $r_{\cal O}$. The general expression for $r_{\cal O}$ is complicated and can be found in \cite{Aalsma:2025lcb}. In the Nariai limit it reduces to $r_{\cal O}=r_b=r_c$.

Adapting coordinates to this observer as
\beq
\bal 
\tilde t &= \sqrt{f(r_{\cal O})} \,t ~, \\
\tilde r &= \frac{r-r_{\cal O}}{\sqrt{f(r_{\cal O})}} ~,
\eal
\eeq
and taking the Nariai limit $(r_b,r_c)\to \ell$ the metric becomes
\beq
\rmd s^2 = -\left(1-\frac{\tilde r}{L^2}\right)\rmd\tilde t^2+\left(1-\frac{\tilde r^2}{L^2}\right)^{-1}\rmd\tilde r^2 + \ell^2 \left(\rmd\theta^2+\sin^2\theta\rmd\phi^2\right) ~.
\eeq
Here, the two-dimensional de Sitter radius $L$ is related to the radius $\ell$ of the two-sphere by
\beq
L = \frac{L_4}{\sqrt{6-L_4^2/\ell^2}} ~.
\eeq
The ratio of radii satisfies $0\leq \ell/L\leq 1$. The limit $\ell/L=0$ corresponds to the maximum charge (ultracold) Nariai black hole that satisfies $r_a=r_b=r_c$ and $\ell/L=1$ to the uncharged Nariai black hole with $r_a=0$ and $r_b=r_c$.

Using the coordinate transformation \eqref{eq:StatToGlobal}, we can write the de Sitter part of the metric in global coordinates
\beq
\rmd s^2 = L^2\left(-\rmd\tau^2+\cosh^2\tau\,\rmd\chi^2\right) + \ell^2 \left(\rmd\theta^2+\sin^2\theta\rmd\phi^2\right) ~.
\eeq

\subsection{From the sphere to de Sitter space}
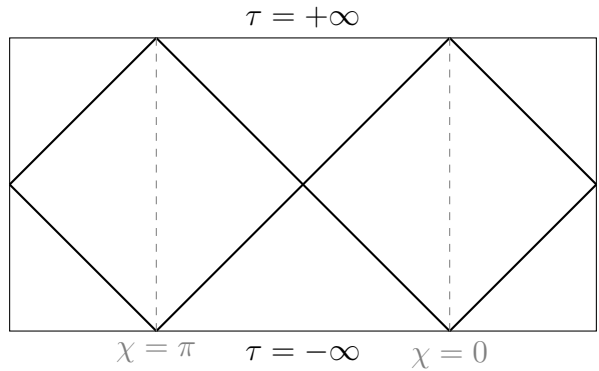
\begin{figure}[t]
\centering
\begin{tikzpicture}[scale=1.3]

\draw [draw] (0,0) rectangle (6,3) node[above] at (3,3) {$\tau=+\infty$}
node[below] at (3,0) {$\tau=-\infty$};
\draw[thick] (0,1.5) -- (1.5,3);
\draw[thick] (0,1.5) -- (1.5,0);
\draw[thick] (1.5,0) -- (4.5,3);
\draw[thick] (1.5,3) -- (4.5,0);
\draw[thick] (4.5,0) -- (6,1.5);
\draw[thick] (4.5,3) -- (6,1.5);

\draw[gray,dashed] (1.5,0) -- (1.5,3) node[pos=0,below]{$\chi=\pi$};
\draw[gray,dashed] (4.5,0) -- (4.5,3) node[pos=0,below]{$\chi=0$};

\end{tikzpicture}
\caption{Penrose diagram of two-dimensional de Sitter space. Unlike higher-dimensional de Sitter space, the two-dimensional Penrose diagram is rectangular. We indicated the north pole at $\chi=0$ and the south pole at $\chi=\pi$.}
\label{fig:dS2Penrose}
\end{figure}

To relate the Euclidean Green's function on $S^2$ to the Lorentzian Green's function on two-dimensional de Sitter space, all we need to do is analytically continue. Sending $\psi\to i\tau+\frac\pi2$ the $S^2$ metric becomes
\beq
\rmd s^2 = L^2\left(-\rmd \tau^2 + \cosh^2\tau\,\rmd\chi^2\right) ~,
\eeq
which we recognize as the two-dimensional de Sitter metric expressed in global coordinates. Its Penrose diagram is given in Figure \ref{fig:dS2Penrose}.
In two dimensions, the Penrose diagram of de Sitter space is rectangular with timelike edges that are identified. The reason for the rectangular shape, in contrast with the square shape in higher dimensions, is that in higher dimensions the Penrose diagram can be viewed as a two-dimensional projection of a timelike and a radial coordinate, where the radial coordinate is restricted to take positive values. In two dimensions, the would-be radial coordinate can also take negative values, extending the Penrose diagram horizontally. In the global coordinates we used, this is reflected by the full range $0\leq \chi\leq 2\pi$ on display in the Penrose diagram.

The exact Green's function in de Sitter space can also be obtained by analytic continuation of \eqref{eq:ExactS2Green}. We then find
\beq \label{eq:dS2GreenExact}
G_{\rm dS_2}(x,y) = \frac{\Gamma(\Delta_+)\Gamma(\Delta_-)}{4\pi}\,_2F_1\left(\Delta_+,\Delta_-,1,\frac{1+Z(x,y)}{2}\right)~,
\eeq
where we found it useful to express the separation between the points $x$ and $y$ in terms of the de Sitter invariant distance
\beq
Z(x,y) = L^{-2}\eta_{AB}X^A(x)X^B(y) ~,
\eeq
where $X^A$ are embedding coordinates, see Appendix \ref{app:coordinates} for more details. The geodesic distance in de Sitter space is related to $Z(x,y)$ via $Z(x,y) = \cos(D(x,y)/L)$. In global coordinates, the geodesic distance is
\beq
D = L\arccos\left(\cos(\chi_2-\chi_1)\cosh\tau_1\cosh\tau_2-\sinh\tau_1\sinh\tau_2\right) ~.
\eeq

We can now easily obtain the geodesic approximation to the Green's function in de Sitter space by analytical continuation, reproducing existing results in \cite{Aalsma:2022eru,Chapman:2022mqd}. In particular, we focus on two special cases where the points are either located at the same pole of de Sitter space or at opposite poles.

For two points at the north pole with $\chi_2-\chi_1=0$ the geodesic distance simplifies to
\beq \label{eq:SamePoleCorr}
\frac DL = i(\tau_2-\tau_1) \qquad (\chi_2-\chi_1 = 0) ~,
\eeq
where the factor of $i$ indicates these points are timelike separated and we assumed $\tau_2-\tau_1\geq 0$. For two points at opposite poles $\chi_2-\chi_1=\pi$ and
\beq
\frac DL = -i(\tau_1+\tau_2) + \pi \qquad (\chi_2-\chi_1 = \pi) ~,
\eeq
where we assumed $\tau_1+\tau_2 \geq 0$.

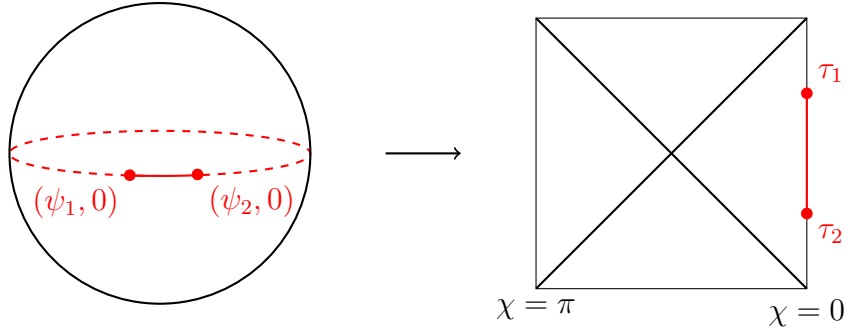
\begin{figure}[t]
\centering
\begin{tikzpicture}

\draw[thick,red]
plot[domain=-100:-80,samples=100]
({2 + 2*cos(\x)}, {0.3*sin(\x)});

\draw[thick,red,dashed]
  plot[domain=-80:270,samples=100]
  ({2 + 2*cos(\x)}, {0.3*sin(\x)});

\draw[thick](2,0) circle (2);

\filldraw[red] (1.6,-.29) circle (2pt) node[anchor=north east]{$(\psi_1,0)$};
\filldraw[red] (2.5,-.28) circle (2pt) node[anchor=north west]{$(\psi_2,0)$};

\draw[thick, ->](5,0) -- (6,0);


\draw [draw] (7,-1.8) rectangle (10.6,1.8) node[below] at (10.6,-1.8) {$\chi=0$}
node[below] at (7,-1.8) {$\chi=\pi$};

\draw[thick] (7,-1.8) -- (10.6,1.8);
\draw[thick] (7,1.8) -- (10.6,-1.8);

\draw[thick,red] (10.6,-.8) -- (10.6,.8);

\filldraw[red] (10.6,-.8) circle (2pt) node[anchor=north west]{$\tau_2$};
\filldraw[red] (10.6,.8) circle (2pt) node[anchor=south west]{$\tau_1$};

\end{tikzpicture}
\caption{Analytical continuation of the geodesic with $\chi_1=\chi_2=0$ on the sphere to two-dimensional de Sitter space. In de Sitter space, the solid geodesic gives the dominant contribution. For compactness, we only display the Penrose diagram for the region $0\leq \chi\leq \pi$.}
\label{fig:SamePole}
\end{figure}

From these expression, we see that we can obtain these two distinct correlators (with $\chi_2-\chi_1=0$ or $\chi_2-\chi_1=\pi$) from analytically continuing the two `fundamental' contributions to the $S^2$ Green's function in \eqref{eq:S2approxInter} in a distinct way. To obtain the de Sitter correlator with distance \eqref{eq:SamePoleCorr} we continue the fundamental geodesic as follows
\beq
D = L(\psi_2-\psi_1) \to i(\tau_2-\tau_1)L ~,
\eeq
After analytically continuing to de Sitter space there is only one (dominant) geodesic that contributes. Hence, we don't need to take into account the geodesic $\bar D$. We visualize this analytic continuation in Figure \ref{fig:SamePole}. 

\begin{figure}[t]
\centering
\includegraphics{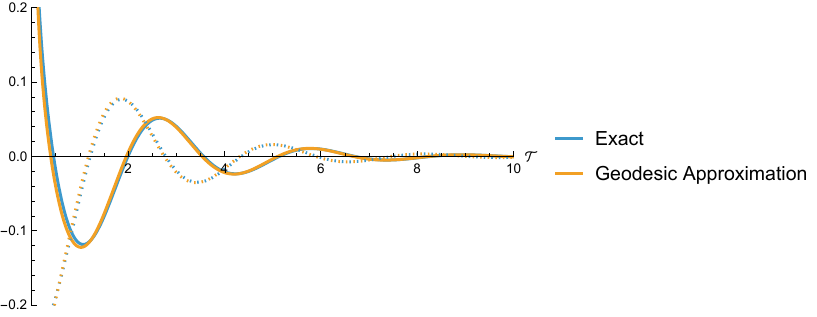}
\caption{Comparison of the exact Green's function \eqref{eq:dS2GreenExact} and the geodesic approximation \eqref{eq:dSsingleGeod}. We took $mL = 2$. The solid lines indicate the real part and the dashed line indicate the imaginary part.}
\label{fig:SingleGeoddS}
\end{figure}

The Green's function is obtained by performing this continuation on \eqref{eq:SingleGeodSphere}
\beq \label{eq:dSsingleGeod}
G_{\rm dS_2}({\cal T}) = \frac{e^{-i(mL{\cal T} +\frac\pi4)}}{\sqrt{8\pi mL\sinh{\cal T}}} ~,
\eeq
where we wrote ${\cal T} = \tau_2-\tau_1$. In Figure \ref{fig:SingleGeoddS} we show that this gives an excellent approximation to the exact result.

\begin{figure}[t]
\centering
\begin{tikzpicture}

\begin{scope}[rotate around={-40:(2.8,0)}]
\draw[thick,red]
plot[domain=65:285,samples=100]
({2.6 - 0.23*sin(\x)}, {-1.44*cos(\x)});
\end{scope}

\begin{scope}[rotate around={-40:(2.8,0)}]
\draw[thick,red,dashed]
plot[domain=280:435,samples=100]
({2.6 - 0.23*sin(\x)}, {-2.45*cos(\x)});
\end{scope}

\filldraw[red] (2.6,-.29) circle (2pt) node[anchor=north west]{$(\psi_1,0)$};
\filldraw[red] (2.08,-.2) circle (2pt) node[anchor=south east]{$(\psi_2,\pi)$};

\draw[thick](2,0) circle (2);

\draw[thick, ->](5,0) -- (6,0);


\draw [draw] (7,-1.8) rectangle (10.6,1.8) node[below] at (10.6,-1.8) {$\chi=0$}
node[below] at (7,-1.8) {$\chi=\pi$};

\draw[thick] (7,-1.8) -- (10.6,1.8);
\draw[thick] (7,1.8) -- (10.6,-1.8);

\draw[thick,red] (10.6,0) -- (10.6,.8);
\draw[thick,red] (7,0) -- (7,.8);
\filldraw[red] (10.6,.8) circle (2pt) node[anchor=south west]{$\tau_1$};
\filldraw[red] (7,.8) circle (2pt) node[anchor=south east]{$\tau_2$};

\draw[thick,color=red] (7,0,0) arc [start angle=-180,end angle=0,x radius=1.8,y radius=0.3];
\draw[thick,color=red,dashed] (7,0,0) arc [start angle=180,end angle=0,x radius=1.8,y radius=0.3];

\end{tikzpicture}
\caption{Analytical continuation of the geodesics with $\chi_2-\chi_1=\pi$ on the sphere to two-dimensional de Sitter space. The Green's function involves contributions from two geodesics that have a conjugate imaginary part. For compactness, we only display the Penrose diagram for the region $0\leq \chi\leq \pi$.}
\label{fig:OppositePole}
\end{figure}
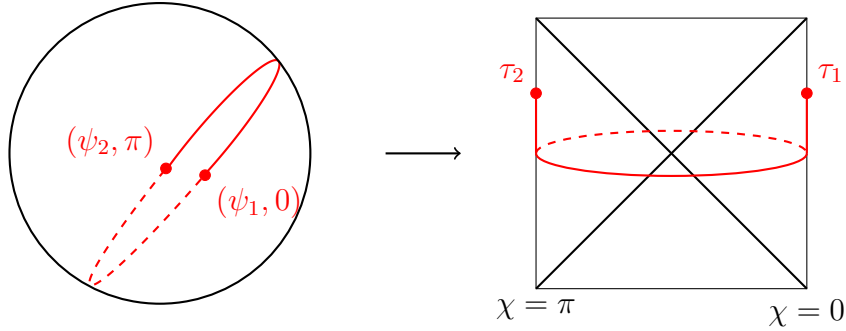

Next, we consider the correlator with points at opposite poles. In this case, both geodesics parametrized by $D$ and $\bar D$ contribute and the de Sitter correlator can be obtained by sending
\beq
\bal \label{eq:OppositePatchContin}
D &= L\left(\pi-(\psi_1+\psi_2) \right) \to L\left(\pi -i(\tau_1+\tau_2) \right) ~, \\
\bar D &= L\left(\pi+(\psi_1+\psi_2)\right) \to L\left(\pi +i(\tau_1+\tau_2) \right)~.
\eal
\eeq
We see that in de Sitter space, these two geodesic are complex in the sense that they involve evolution in Lorentzian (real) time in addition to a complex piece equal to $\pm \pi$. The two geodesics that contribute are each others complex conjugate. This analytic continuation is displayed geometrically in Figure \ref{fig:OppositePole}.

\begin{figure}[t]
\centering
\includegraphics{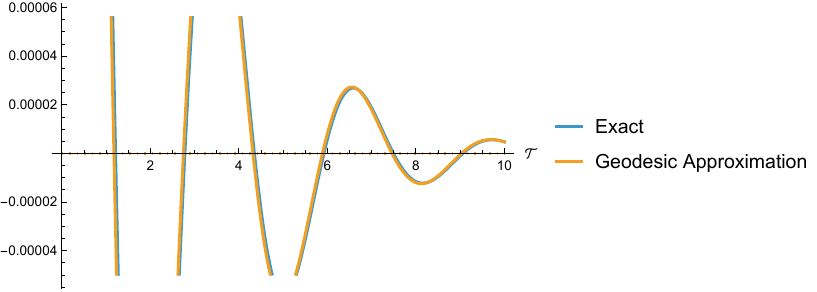}
\caption{Comparison of the exact Green's function \eqref{eq:dS2GreenExact} in two-dimensional de Sitter space and the geodesic approximation \eqref{eq:DoubleGeoddS}. We took $mL = 2$.}
\label{fig:DoubleGeoddS}
\end{figure}

The correlator for points in opposite static patches is therefore obtained by analytically continuing both contributions, i.e. 
\beq \label{eq:DoubleGeoddS}
G_{\rm dS_2}({\cal T}) = \frac{e^{-mL\pi}}{\sqrt{4\pi mL\sinh{\cal T}}}\left(\cos(mL{\cal T})+\sin(mL{\cal T})\right) ~.
\eeq
Here ${\cal T}=\tau_1+\tau_2$. In Figure \ref{fig:DoubleGeoddS} we show that \eqref{eq:DoubleGeoddS} gives an excellent approximation to \eqref{eq:dS2GreenExact}.

\subsection{From products of spheres to the Nariai geometry}
\begin{figure}[t]
\centering
\begin{tikzpicture}

\begin{scope}[scale=1.25]
\draw [draw] (0,0) rectangle (6,3) node[above] at (3,3) {$\tau=+\infty$}
node[below] at (3,0) {$\tau=-\infty$};
\draw[thick] (0,1.5) -- (1.5,3);
\draw[thick] (0,1.5) -- (1.5,0);
\draw[thick] (1.5,0) -- (4.5,3);
\draw[thick] (1.5,3) -- (4.5,0);
\draw[thick] (4.5,0) -- (6,1.5);
\draw[thick] (4.5,3) -- (6,1.5);

\draw[gray,dashed] (1.5,0) -- (1.5,3) node[pos=0,below]{$\chi=\pi$};
\draw[gray,dashed] (4.5,0) -- (4.5,3) node[pos=0,below]{$\chi=0$};

\filldraw[blue] (1.5,2.2) circle (2pt) node[anchor=south west]{$\tau_1$};
\filldraw[blue] (4.5,2.2) circle (2pt) node[anchor=south east]{$\tau_2$};

\draw[thick,blue] (1.5,1.5,0) arc [start angle=-180,end angle=0,x radius=1.5,y radius=0.3];
\draw[thick,blue,dashed] (1.5,1.5,0) arc [start angle=180,end angle=0,x radius=1.5,y radius=0.3];

\draw[thick,blue] (1.5,1.5) -- (1.5,2.2);
\draw[thick,blue] (4.5,1.5) -- (4.5,2.2);

\draw[thick, orange] (5.0, 1.9) rectangle (5.2, 2.1);
\end{scope}


\begin{scope}[shift={(8.4,3.5)},scale=.6]

\draw[thick, orange] (-0.2,-2.2) rectangle (4.2,2.2);

\draw[thick,red]
plot[domain=-100:-80,samples=100]
({2 + 2*cos(\x)}, {0.3*sin(\x)});

\draw[thick,red,dashed]
  plot[domain=-80:270,samples=100]
  ({2 + 2*cos(\x)}, {0.3*sin(\x)});

\draw[thick](2,0) circle (2);

\filldraw[red] (1.6,-.29) circle (2pt) node[scale=.6,anchor=north east]{$(\theta_1,0)$};
\filldraw[red] (2.5,-.28) circle (2pt) node[scale=.6,anchor=north west]{$(\theta_2,0)$};

\end{scope}

\draw[thick,orange, ->] (6.6, 2.6) .. controls (7.3,3.1) .. (8.1, 3.3);

\end{tikzpicture}
\caption{Visualization of the four geodesics that contribute to the Green's function \eqref{eq:NariaiGeod} in the Nariai geometry. Each point in the Penrose diagram corresponds to a sphere, which we highlighted by the orange box.}
\label{fig:NariaiGeo}
\end{figure}
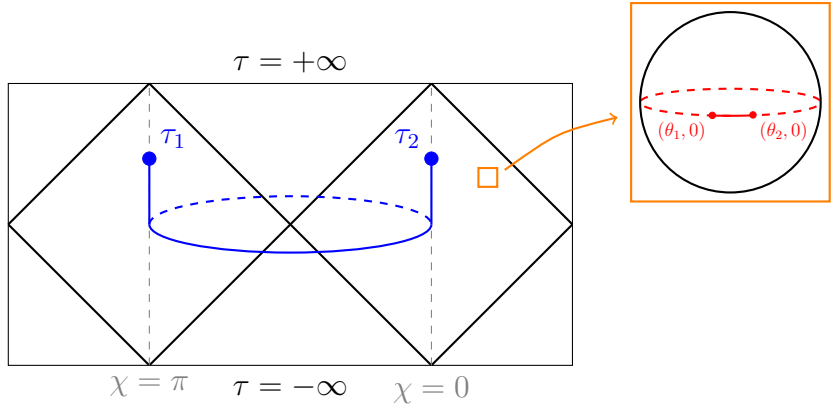

\begin{figure}[t]
\centering
\includegraphics{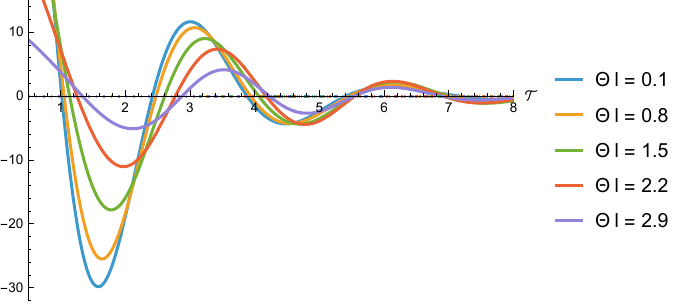}
\caption{Geodesic approximation to the Green's function in the Nariai geometry, given by \eqref{eq:NariaiGeod}, for different values of $\Theta \ell$. We took $L = 1$ and $m=2$.}
\label{fig:NariaiGeodPlot}
\end{figure}

We can now perform the analytic continuation necessary to obtain the geodesic approximation to the Green's function in the Nariai geometry. Since we're mostly interested in complex geodesics, we focus on points that are located in opposite static patches. Using the analytical continuation applied to the different contributions \eqref{eq:S2S2terms} of the $S^2\times S^2$ Green's function, we continue one of the spheres to de Sitter space. The geometry is then given by
\beq
\rmd s^2 = L^2\left(-\rmd\tau^2+\cosh^2\tau\,\rmd\chi^2\right) + \ell^2 \left(\rmd\theta^2+\sin^2\theta\rmd\phi^2\right) ~.
\eeq
We pick points such that on the de Sitter part of the geometry, they are located in opposite static patches with a separation ${\cal T} = \tau_1+\tau_2$. On the two-sphere, we separate them by angle $\Theta = \theta_2-\theta_1$ and keep the separation in $\phi$ zero. The Green's function is then given by
\beq \label{eq:NariaiGeod}
G_{{\rm dS}^2\times S^2}({\cal T},\Theta) = \left({\cal G}_{1,2}+{\cal G}_{\bar 1,2}+e^{i\frac{\pi}{2}}{\cal G}_{1,\bar2}+e^{-i\frac\pi 2}{\cal G}_{\bar1,\bar2}\right) ~,
\eeq
where 
\beq
{\cal G}_{i,j} = \sqrt{\frac{m(D_i/\ell_i)(D_j/\ell_j)}{32\pi^3\sin(D_i/\ell_i)\sin(D_j/\ell_j)(D_i^2+D_j^2)^{3/2}}}e^{-m\sqrt{D_i^2+D_j^2}} ~.
\eeq
As before, $i,j$ run over 1,2 and the bar denotes conjugate geodesics. Again, we note that the phases in front of the last two contributions are such that the result for the Green's function does not diverge in the limit $\Theta\to\pi$. Explicitly, the different contributions are given by
\beq
\bal
D_1 &= L(\pi -i{\cal T}) ~,\\
\bar D_1 &= L(\pi +i{\cal T}) ~,\\
D_2 &= \ell\Theta ~, \\
\bar D_2 &= \ell(2\pi - \Theta) ~.
\eal
\eeq
This Green's function consists of a sum of four different geodesics, two for both the de Sitter and the sphere factor. The contributions are visualized in Figure \ref{fig:NariaiGeo}.
In this figure, we connected the two points on the de Sitter factor (in blue) through the region $0\leq \chi\leq \pi$. However, we could have also chosen to display these geodesics as connecting through the region $\pi \leq \chi\leq 2\pi$. The resulting Green's function is the same and we display it for different values of $\Theta$ in Figure \ref{fig:NariaiGeodPlot}.

Physically, the difference between these two routes is whether the (complex) geodesic is moving trough the black hole or cosmological horizon \cite{Faruk:2023uzs}. In the Nariai limit, however, the distinction between these two horizons disappears, This fact, which has long been appreciated \cite{Susskind:2021omt}, results in  indistinguishable Green's functions. Away from the Nariai limit (for a non maximal-size black hole) this degeneracy should be lifted.

\section{Discussion} \label{sec:discussion}
In this paper, we have studied the behavior of the Green's function and complex geodesics in the Nariai geometry. Utilizing heat kernel methods, we constructed the exact Green's function on a product of spheres and took a large-mass limit to obtain a geodesic approximation. For each of the spheres, the leading contribution for an arbitrary separation between points consists of the two distinct geodesics that connect two points on the sphere. Importantly, the indirect geodesic (the one that does not have the shortest path) picks up a phase that is crucial to ensure the resulting approximate Green's function is real and does not contain any spurious singularities.

By analytically continuing one of the spheres, we obtain the geodesic approximation to the Green's function in the Nariai geometry, which is given by four contributions that (in the Euclidean picture) correspond to the two distinct geodesics on each of the spheres. The result displays a degeneracy in the sense that in the near-horizon geometry, two points in opposite static patches can be connected by (complex) geodesics either through the black hole or cosmological horizon. The result for the Green's function is identical in either situation.

This is not surprising; from the perspective of the near-horizon geometry the black hole and cosmological horizon are identical and in this limit the Penrose diagram no longer displays the black hole singularity. We expect this degeneracy to be lifted by quantum corrections that induce a decay of the Nariai black hole towards a smaller black hole. Since this changes the geometry in a drastic way\footnote{In particular, a geometry with a black hole singularity allows two points on opposite sides of the black hole horizon to be connected by a real geodesic \cite{Fidkowski:2003nf}.}, it would be quite interesting to see how this modifies our result. One option to do so is to consider a near-Nariai Jackiw-Teitelboim model. By introducing a dilatonic degree of freedom it becomes possible to distinguish the black hole interior from the cosmological exterior by the value of the dilaton, without completely destroying the symmetries of the near-horizon geometry. Furthermore, such a model could also be used to study other types of extremal limits in de Sitter space, such as the one corresponding to the cold black hole, which as an AdS$_2\times S^2$ near-horizon geometry.

Other interesting future directions include studying geodesics of different fields. In particular, looking at charged or spin degrees of freedom presumably allows us to distinguish different Nariai geometries. In addition, our results can likely be used to study quasinormal modes (recent studies include \cite{Faruk:2025bed}), as those can be extracted from the Green's function. This seems especially timely in light of recent discussion about one-loop corrections to the Euclidean partition of black holes in de Sitter space \cite{Maulik:2025phe,Turiaci:2025xwi,Blacker:2025zca,Arnaudo:2025btb,Aalsma:2025lcb}, since such corrections are closely related to quasinormal modes \cite{Kapec:2023ruw}.

\subsection*{Acknowledgments}
LA is supported by the President's Postdoctoral Fellowship Program at the University of Minnesota. MMF acknowledges funding support from the
Double First Class Postdoctoral Fellowship and Prof. Alexey Koshelev’s start-up
fund (2023F0201-000-01).
\appendix

\section{Identities}
We encountered the following sum over Legendre polynomials
\beq
I = \sum_{l=0}^\infty \frac{2l+1}{l(l+1)+m^2\ell^2}P_l(p) ~.
\eeq
This sum can be rewritten in terms of a hypergeometric function, by first using the identity \cite{DLMF}
\beq
P_\nu(-x) = \frac{\sin\pi\nu}{\pi}\sum_{n=0}^\infty \frac{2n+1}{(\nu-n)(\nu+n+1)}P_n(x) ~.
\eeq
We subsequently use the following representation of the hypergeometric function \cite{WikiHypergeometricFunction}
\beq \label{eq:LegendreAsHyper}
\,_2F_1(\nu,1-\nu,1,z) = P_{-\nu}(1-2z) ~,
\eeq
to write
\beq \label{eq:HyperLegendre}
I = \frac{\pi}{\sin(\pi\Delta_+)}\,_2F_1\left(\Delta_+,\Delta_-,1,\frac{1+p}{2}\right) ~.
\eeq
Here we identified $\nu = \Delta_+$ and
\beq
p = 2z-1 ~, \quad \Delta_\pm = \frac12\left(1\pm \sqrt{1-4m^2\ell^2}\right) ~.
\eeq
Another useful identity is the asymptotic expansion of the Legendre polynomials with an imaginary argument. That is, $P_{-\frac12+ix}(\cos\theta)$ in the limit $x\to \infty$ becomes \cite{DLMF}
\beq \label{eq:asympLegendre}
P_{-\frac12-ix}(\cos\theta) = \sqrt{\frac\theta{\sin\theta}}\,I_0(x\theta)\left(1+{\cal O}(1/x)\right) ~.
\eeq

\section{Coordinates} \label{app:coordinates}
Here, we discuss the coordinates that we use to describe the $S^2$ and dS$_2$ geometries.

\subsubsection*{Two-sphere}
We can describe an $S^2$ by embedding it into three-dimensional Euclidean flat space as
\beq
\delta_{AB}X^AX^B = \ell^2 ~.
\eeq
Here $\delta_{AB}$ is the three-dimensional flat Euclidean metric and we define coordinates
\beq
\bal
X^0 &= \ell \sin\theta\cos\phi ~,\\
X^1 &= \ell\sin\theta\sin\phi ~,\\
X^2 &= \ell\cos\phi ~.
\eal
\eeq
The line element is given by
\beq
\rmd s^2 = \delta_{AB}\rmd X^A\rmd X^B= \ell^2\left(\rmd\theta^2+\sin^2\theta\rmd\phi^2\right) ~.
\eeq
Now consider two points $x=(\theta_1,\phi_1)$ and $y=(\theta_2,\phi_2)$. The geodesic distance $D(x,y)$ between these two points is given by
\beq
\bal
\cos\left(\frac{D(x,y)}{\ell}\right) &= \ell^{-2}\delta_{AB}X^A(x)X^B(y) \\
&=\cos\theta_1\cos\theta_2+\cos(\phi_2-\phi_1)\sin\theta_1\sin\theta_2 ~.
\eal
\eeq

\subsubsection*{De Sitter space: global coordinates}
Similarly, the dS$_2$ geometry can be described by embedding it into three-dimensional Minkowski space as
\beq
\eta_{AB}X^AX^B = L^2 ~,
\eeq
where $\eta_{AB}$ is the metric of three-dimensional Minkowkski space. We define global coordinates as
\beq
\bal
X^0 &= L \cosh\tau\cos\chi ~,\\
X^1 &= L\cosh\tau\sin\chi ~,\\
X^2 &= L\cos\chi ~.
\eal
\eeq
We see that these coordinates are obtained by analytical continuation of the $S^2$ coordinates by sending $(\theta,\phi)\to (i\tau+\frac\pi 2,\chi)$. The line element is
\beq
\rmd s^2 = \eta_{AB}\rmd X^A\rmd X^B = L^2\left(\rmd\tau^2+\cosh^2\tau\rmd\chi^2\right) ~.
\eeq
For two points $x=(\tau_1,\chi_1)$ and $y=(\tau_2,\chi_2)$ the geodesic distance is
\beq
\bal
\cos\left(\frac{D(x,y)}{\ell}\right) &= L^{-2}\eta_{AB}Y^A(x)Y^B(y) \\
&= \cosh\tau_1\cosh\tau_2\cos(\chi_2-\chi_1) - \sinh\tau_1\sinh\tau_2 ~.
\eal
\eeq

\subsubsection*{De Sitter space: static coordinates}
Lastly, we introduce static coordinates on de Sitter space. The appear naturally in the near-horizon Nariai limit of black holes in de Sitter space. The embedding coordinates are given by
\beq
\bal
X^0 &= \sqrt{L^2-\tilde r^2}\sinh(\tilde t/L) ~,\\
X^1 &= \sqrt{L^2-\tilde r^2}\cosh(\tilde t/L) ~, \\
X^2 &= \tilde r ~.
\eal
\eeq
The metric is given by
\beq
\rmd s^2 = \eta_{AB}\rmd X^A\rmd X^B = \left(1-\frac{\tilde r^2}{L^2}\right)\rmd\tilde t^2 + \left(1-\frac{\tilde r^2}{L^2}\right)^{-1}\rmd \tilde r^2 ~.
\eeq
These coordinates can be related to global coordinates by the following coordinate transformation
\beq \label{eq:StatToGlobal}
t = L\,\text{arctanh}\left(\tanh\tau\sec\chi\right)~, \qquad r = L\cosh\tau\sin\chi~.
\eeq

\bibliographystyle{utphys}
\bibliography{refs}

\end{document}